\documentstyle[12pt,epsf]{l-aa}

\def\spose#1{\hbox to 0pt{#1\hss}}
\def\simlt{\mathrel{\spose{\lower 3pt\hbox{$\mathchar"218$}}
     \raise 2.0pt\hbox{$\mathchar"13C$}}}
\def\simgt{\mathrel{\spose{\lower 3pt\hbox{$\mathchar"218$}}
     \raise 2.0pt\hbox{$\mathchar"13E$}}}
\def\simpropto{\mathrel{\spose{\lower 3pt\hbox{$\mathchar"218$}}
     \raise 2.0pt\hbox{$\propto$}}}

\catcode`@=11
\def\eqalign#1{\null\,\vcenter{\openup\jot\m@th
  \ialign{\strut\hfil$\displaystyle{##}$&$\displaystyle{{}##}$\hfil
      \crcr#1\crcr}}\,}
\def\eqalignleft#1{\null\,\vcenter{\openup\jot\m@th
  \ialign{\strut$\displaystyle{##}$\hfil&$\displaystyle{{}##}$\hfil
      \crcr#1\crcr}}\,}
\catcode`@=12 

\begin{document}
\setcounter{footnote}{1}

\thesaurus{01      
      (08.14.1; 
       08.09.2: Her X-1)} 
\title{On some features of free precession of a triaxial body: the case of
Her X-1}

\author{N.I.Shakura\inst{1},
K.A.~Postnov\inst{2}
\and M.E.~Prokhorov\inst{1}}

\institute{Sternberg Astronomical Institute, Moscow University,
                119899 Moscow, Russia
\and
Faculty of Physics, Moscow State University,
                119899 Moscow, Russia
}
\date{Received ... 1997, accepted ..., 1997}
\maketitle
\markboth{N.Shakura, K.Postnov \& M.Prokhorov. Triaxial free precession}{ ...}

\begin{abstract}
We show that the free precession of a triaxial body 
can naturally explain the  anomalously rapid change of 
the X-ray  pulse profile of Her X-1 
observed by the HEAO-1 in September 1978
without requiring a large change in the moment of inertia.


\end{abstract}

Hercules X-1 is the most famous and well studied X-ray pulsar containing an
accreting neutron star with a spin period of $P_{ns}=1.24$ s in a circular
orbit around a $2M_\odot$ main-sequence star. The orbital period of the
binary system is 1.7 days. Discovered in 1972 (Tananbaum et al. 1972), it
nevertheless has not been completely understood until now. This mostly
concerns the origin of its long-term 35-day X-ray periodicity, which has
broadly been discussed in the literature. The possible reason for this
long-term period was suggested to be either (1) the neutron star free
precession (Brecher 1972) or (2) the precession of a tilted accretion disk
controlled by the outer parts (i.e. by the precession of HZ Her (Roberts
1974) or by some intrinsic reasons (Boynton et al. 1980)). Notice that 
both the neutron star free precession and a complex precessing
motion of the accretion disc may in fact simultaneously 
operate in Her X-1/HZ Her binary
system (Shakura et al. 1997, in preparation).  

A strong evidence favouring the free precession model was found by
Tr\"umper et al. (1986) in the EXOSAT observations of the X-ray pulse
profile phase and shape changing over the 35-day period. In contrast, Soong
et al. (1987) claimed that their observations of Her X-1 by HEAO-1 X-ray
satellite in 1978 do not support this model. In September 1978, they
observed an unusually short high-on state of Her X-1 (the X-ray emission
faded down very rapidly during 7 days instead of 10, and the pulse profile
shape changed over 20 hrs, as contrasted to about 17 days for the EXOSAT
observations), which, if interpreted in terms of the free precession model,
would correspond, as the authors claim, to a very large change in the moment
of inertia of the neutron star body 
corresponding to an oblateness of $\sim 8\times
10^{-6}$. Such a large moment of inertia changing would lead in turn to
the pulse period change by an order of magnitude higher than was actually 
observed during this period (i.e. in  August-September 1978) 
$8\times 10^{-7}$ s.

The purpose of this Letter is to show that in fact the free precession model
cannot be so easily rejected if one considers the possible {\it triaxiality}
of the neutron star body. Then the observed episode of an unusually rapid
X-ray pulse shape change in Her X-1 can naturally be explained by a sudden
small deviation of the moment of inertia along one axis without changing the
characteristic period of the free precession (i.e. conserving the gross
oblateness of the body). The magnetic pole simply starts moving
non-uniformly along a non-planar trajectory which apparently manifests
itself as the rapid change in the X-ray pulse shape because the X-ray beam
goes rapidly down to the rotational equator of the neutron star (then the
observer sees two poles producing two equal X-ray pulses over one spin
period, as was observed by the EXOSAT during the low-on state of Her X-1
in March 1984)
and travels the way it usually takes a 17-day interval in a much shorter
time of $\sim 1$ day. Of course, the total moment of inertia of the neutron
star remains practically unchanged and, subject to the angular momentum
conservation, no appreciable X-ray pulse period change should be observed.

\begin{figure*} 
\hbox to\hsize{\hss 
\epsfysize=0.4\hsize\epsfbox{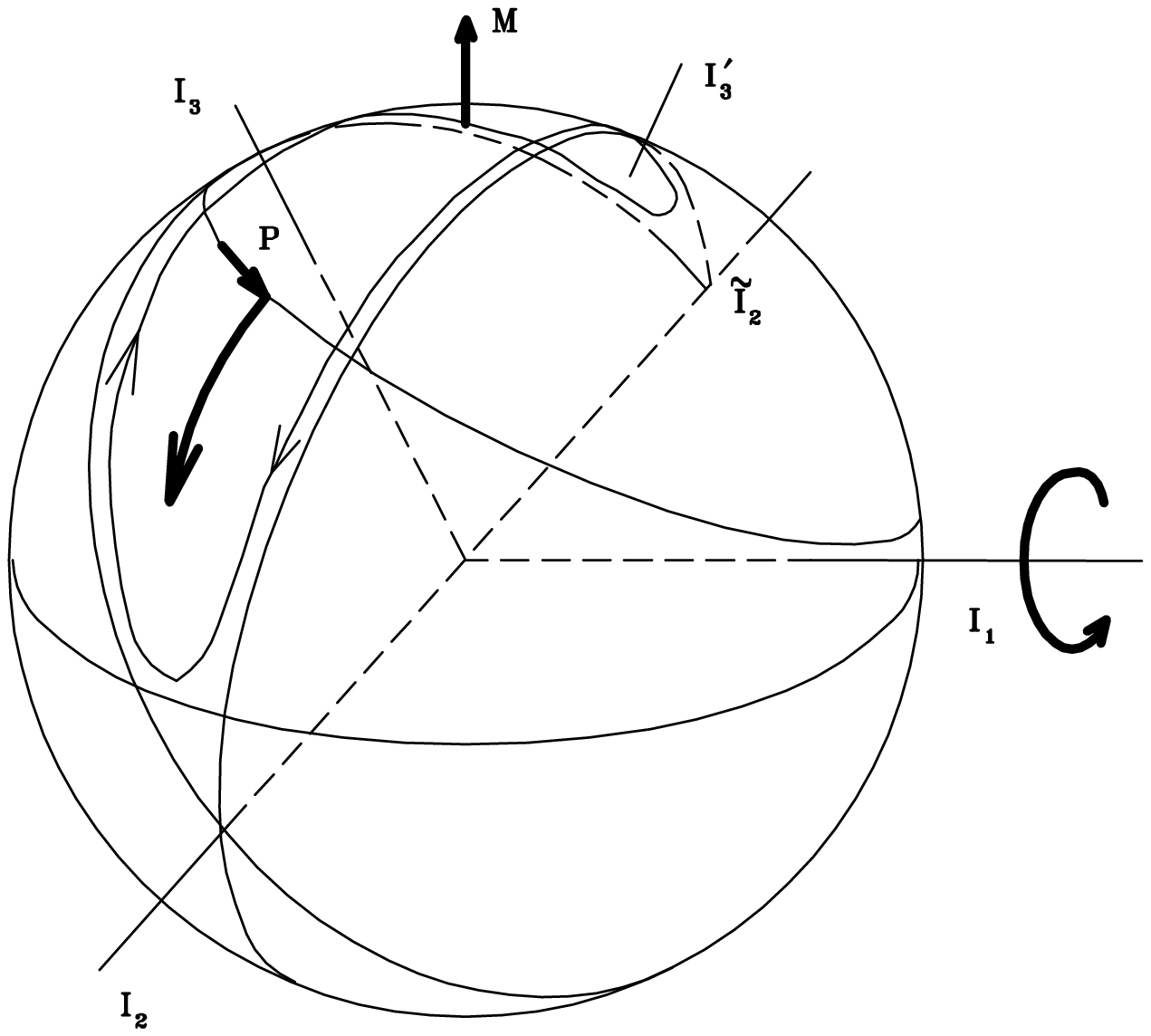}
\hss 
\epsfysize=0.4\hsize\epsfbox{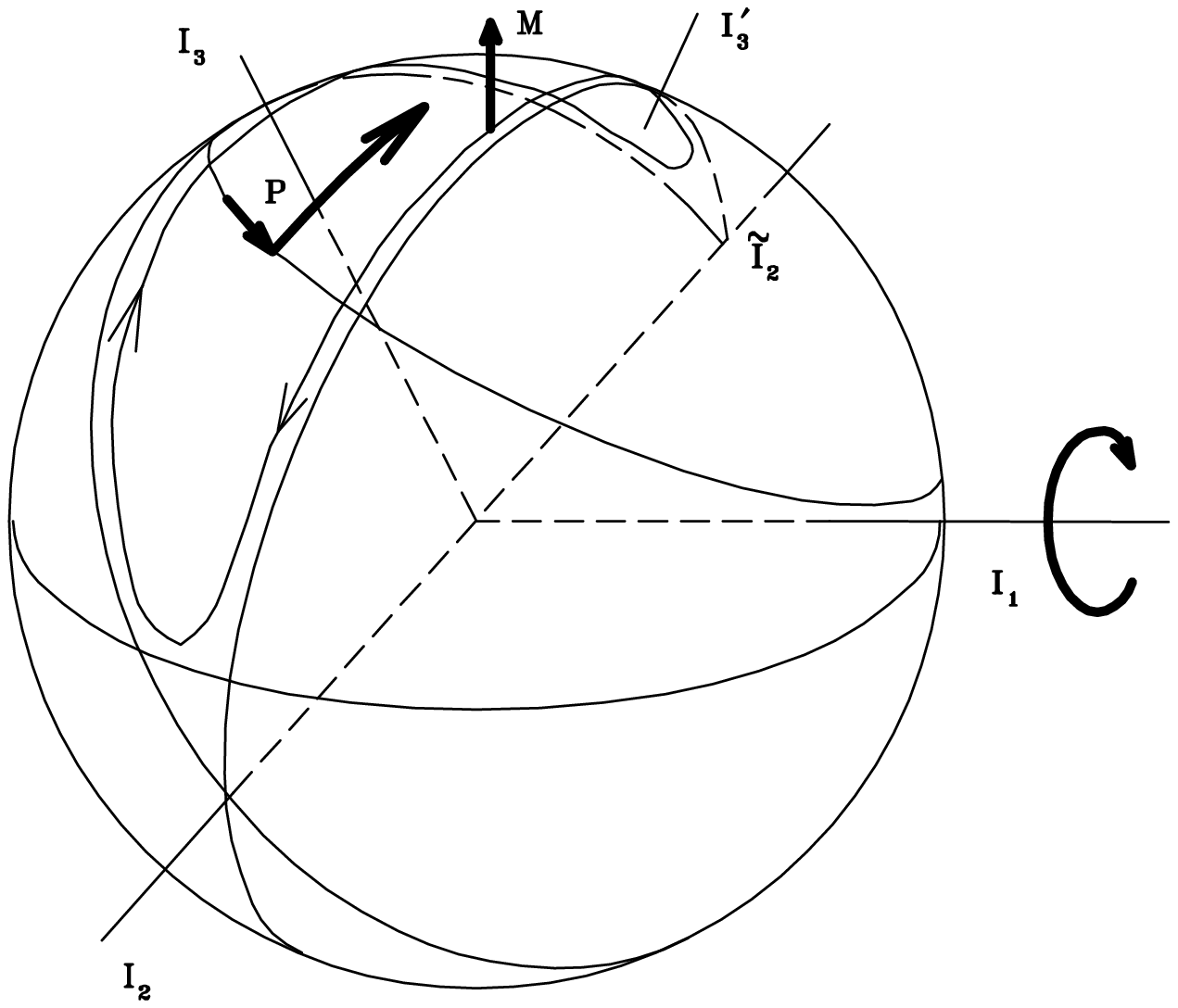} 
\hss} 
\caption{A schematic view
of the neutron star body. $\bf M$ is the angular momentum vector. 
The case of
axisymmetric free precession: $I_3'>I_2'=I_1'$, the magnetic pole moves
along a plane trajectory; the small thick arrow shows the way the pole
passes in 1-day time interval. The case of the triaxial free precession:
$I_3\simgt I_2 > I_1$, two separatrices appear crossing at $I_2$ and $\tilde
I_2$; a non-planar trajectory of $\bf M$ relative to the new axes of inertia
is shown with the thin arrows indicating the direction of the angular
momentum motion. In the left panel, the case when $\bf M$ goes toward
$\tilde I_2$ is shown, i.e. the neutron star body turns anti-clockwise around
an axis close to $I_1$ . In the right panel, $\bf M$ moves toward $I_2$ and
the star turns clockwise around $I_1$. The long thick arrow indicates the
rapid motion of the magnetic pole $P$ toward the rotational equator.}
\label{fig1} 
\end{figure*}

Consider first the more familiar case of an axially symmetric body with
$I_3'>I_2'=I_1'$ rotating around the angular momentum $\bf M$ (see Fig. 1).
Then in the rotating frame the trajectories the magnetic pole of the neutron
star moves along represent plane circles on the neutron star surface with
the center at the largest moment of inertia ($I_3'$). This situation
corresponds to a "normal" free precession in Her X-1 and the magnetic pole
$P$ uniformly goes along such a circle passing in one day a path marked by
the short thick arrow. The precession period is simply $P_{pr}\approx
P_{ns}I_{||}/(\cos b (I_\perp-I_{||}))>>P_{ns}$, where $I_{||}$ and
$I_\perp$ are the components of the moment of inertia parallel and normal to
the total angular momentum and $b$ is the angle between the largest moment
of inertia ($I_3'$ in our case) and the angular momentum.

Let now the body of the neutron star experience some quake resulting in a
practically instantaneous change in all moments of inertia with $I_3 > I_2>
I_1$ (Fig. 1). As is well known (see Landau and Lifshits 1965), in this case
the motion of the angular momentum vector relative to the axes of inertia
becomes more complicated: two families of non-planar trajectories appear
isolated by two separatrices passing through $I_2$, one around $I_3$,
another around $I_1$. The motion along a trajectory around the maximal
moment of inertia ($I_3$) becomes very nonuniform (see Fig. 2): the closer
the trajectory to the separatrix, the more nonuniform the motion along it
is. In Fig. 2 we show how the angle between the angular momentum and the
magnetic pole $\theta$ changes with time over one precession period (see the
Appendix for more detail). The angular momentum rapidly passes most part of
the trajectory and slows down its motion near the turning point close to the
points $I_2$ and $\tilde I_2$ (in the limiting case when the pole goes
exactly along the separatrix, it would stay infinitely long at the
separatrix crossing points $I_2$ and $\tilde I_2$, being in the state of
indifferent equilibrium). In the middle panel of Fig. 2 we also
reproduce the phase change of $\cos\theta$ in the axisymmetrical
case with the angles taken from Tr\"umper et al. (1986)(the thick sinusoid). 
Clearly, 
the quake must have taken place somewhere between $\Psi_{35}=0$
and $\Psi_{35}=0.1$, which indeed corresponds to the observations
of Soong et al. (1987)\footnote{We remind that conventionally 
the precession phase $\Psi_{35}=0$ corresponds to the maximum 
X-ray flux so that the main X-ray turn-on starts at $\Psi_{35}=-0.15$}.

\begin{figure*}
\hbox to\hsize{\hss
\epsfysize=0.32\hsize\epsfbox{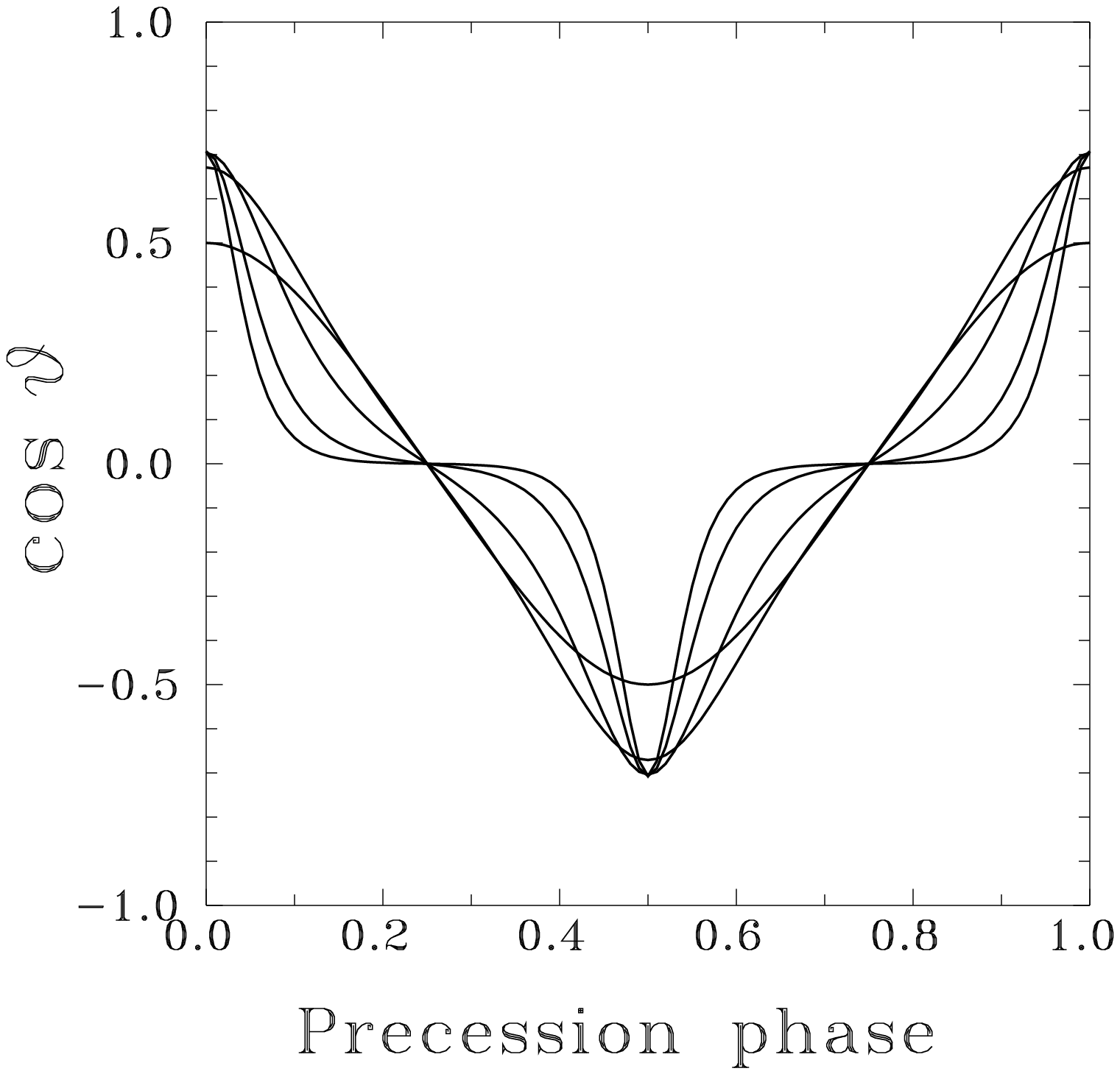}
\hss
\epsfysize=0.32\hsize\epsfbox{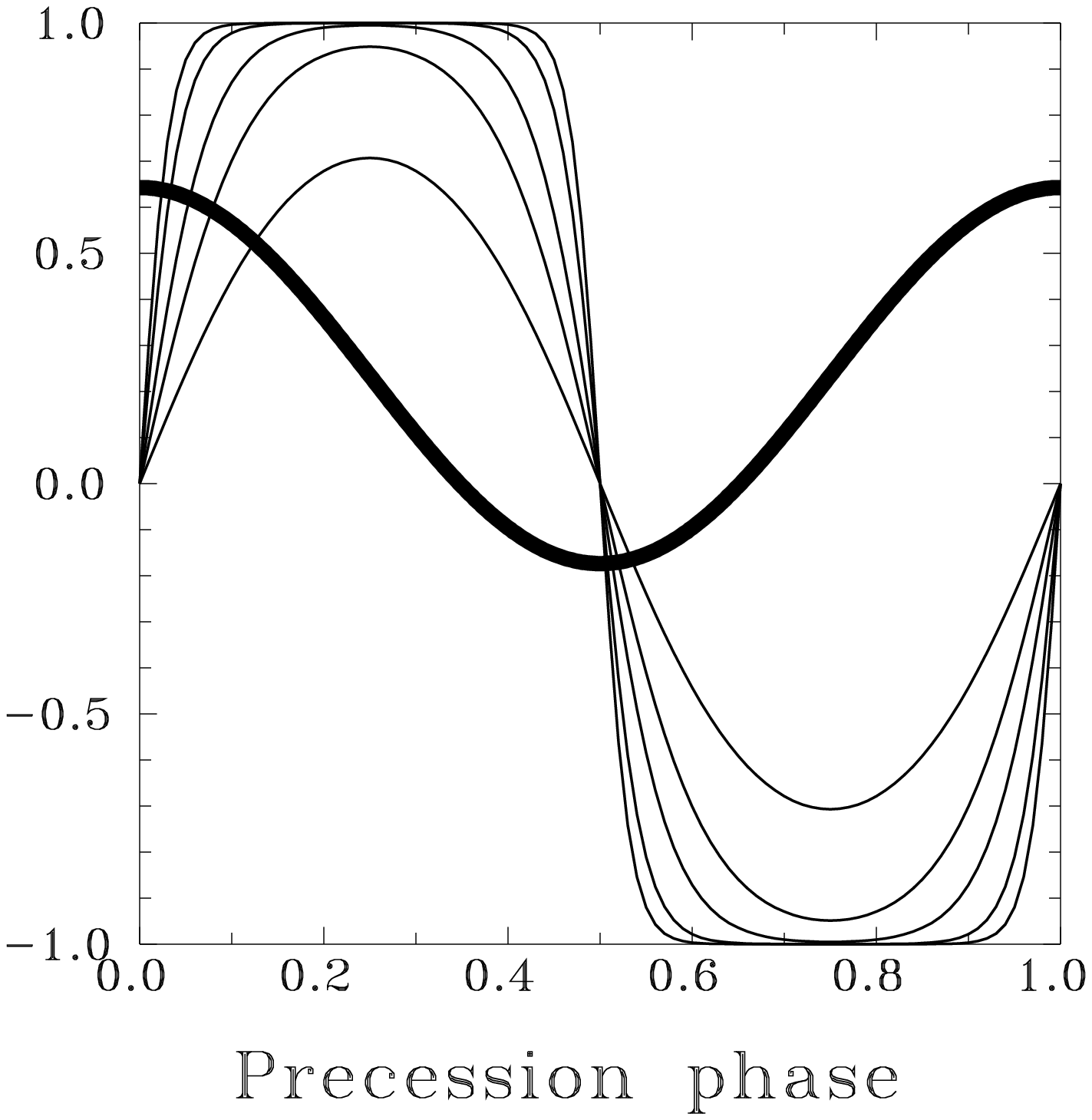}
\hss
\epsfysize=0.32\hsize\epsfbox{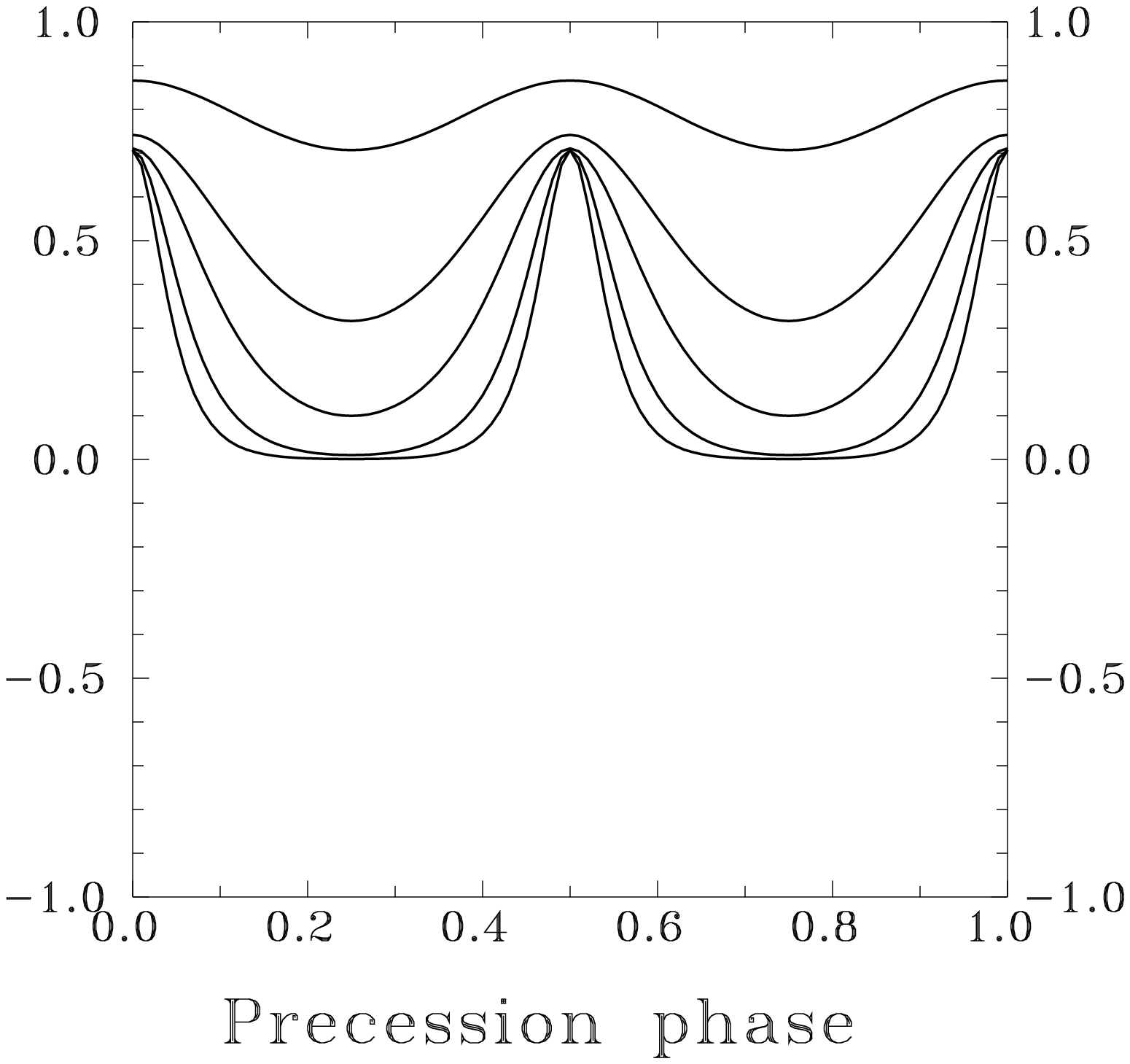}
\hss}
\caption{The dependence of the angle $\theta$ between the magnetic pole P
and the vector of angular momentum $\bf M$ on the precession phase
$\Psi_{35}$
in the
case of the triaxial precession. The relative differences in moments of inertia 
are both $10^{-6}$. The magnetic pole position is close to $I_1$ (left panel),
$I_2$ (middle panel), and $I_3$ (right panel). The five curves in each 
figure are shown for the trajectories 
around $I_3$ (see Fig. 1) characterized by different 
maximal angles $\chi_{max}$ between $\bf M$ and  the axis $I_3$:
$\cot \chi_{max}=1,1/3,1/10,1/100,1/1000$. The closer the trajectory to the
separatrices, the more nonuniform the motion of $\bf M$ along it is.
The thick sinusoid in the middle panel depicts the phase 
behaviour of $\cos \theta$ for axisymmetric precessional motion 
with the angles taken from Tr\"umper et al. (1996):
$\cos\theta=\cos (25^o) \cos (75^o)+\sin (25^o)\sin (75^o)\cos\Psi_{35}$.
The quake must have taken place close to $\Psi_{35}=0.05$
}
\label{fig2}
\end{figure*}

In the triaxial case, the angular momentum vector can move in two opposite
directions depending on at which part of the trajectory the quake happened
(left and right schemes in Fig. 1). Accordingly, in the rotating frame with
the z-axis along $\bf M$, the magnetic pole will rapidly move downward (left
part of Fig. 1) or upward (right part of Fig. 1) since the neutron star body
turns around some axis (close to $I_1$ in Fig. 1). Requiring that the
magnetic pole lies near the rotational equator shortly after the quake (in
order to make it possible to observe an X-ray pulse with two equal peaks),
it should be located near the circle passing through $I_3$ and $I_1$ axes of
inertia (as the angular momentum vector "freezes" near the axis $I_2$).

The long thick arrow in Fig. 1 illustrates the way the magnetic pole now
passes over one day. In Her X-1, the transition from one trajectory to
another occurs between September 22 and 23, 1978, which explains the
apparent 10-fold increase in the free precession rate. After that the
moments of inertia relaxes to their "usual" values and the magnetic pole
returns to another planar "axially-symmetric" trajectory lying not far from
the old one (because the angular momentum spent most time near the
separatrix "crotch").

During the triaxial motion described above the precession period should not
change appreciably since the gross difference in the parallel and
perpendicular moments of inertia remains practically the same. After the
body of the neutron star has returned into its axisymmetric form, it should
be recognized that the X-ray pulse should generally be phase-shifted. This
effect can in principle be detected by accurate timing of X-ray pulses
in different 35-day cycles.

In the free precession model for Her X-1 the vector of the neutron star
angular momentum should be inclined to the line of sight by an angle of
$\sim -40$ degrees tilted away from the observer (see Tr\"umper et al. 1986
for a detailed discussion of all relevant angles in the case of the axially
symmetric free precession). That the angular momentum of the neutron star
proves to be tilted with respect to the orbital angular momentum is
naturally explained in the framework of the free precession model because
the torques applied to a strongly magnetized rotating neutron star by the
accretion disk change the sign for some critical inclination ($\sim 55$
degrees) of the magnetic dipole axis to the neutron star spin axis (Lipunov
1992).

Note to conclude that free precession of neutron stars makes them 
an interesting potential source of gravitational radiation
(Jones 1998), and the confirmation of the free precession model 
for Her X-1 should stimulate such studies. 

\vskip\baselineskip

The work was partially supported by Russian Fund for Basic Research
through Grant No 95-02-06053-a, by the INTAS Grant No 93-3364 and
by the Russian Ministry of Science NTP ``Astronomija'', project 1.4.4.1. 
The authors thank Dr. A.N.Rakhmanov for help in drawing the figures
and the anonymous referee for valuable notes.

\section*{Appendix}

For $I_3>I_2>I_1$, given the total energy 
$$
2E=I_1\Omega_1^2+I_2\Omega_2^2+I_3\Omega_3^2\,,
$$
and angular momentum 
$$
M^2=I_1^2\Omega_1^2+I_2^2\Omega_2^2+I_3^2\Omega_3^2\,,
$$
the motion of the angular momentum vector in the coordinate 
system related to axes of inertia in the rotating frame 
is given by the following system of 
equations (Landau and Lifshits 1965):
$$
\eqalign{
&\Omega_1=\sqrt{\frac{2EI_3-M^2}{I_1(I_3-I_1)}}
\hbox{cn}\,\tau\cr
&\Omega_2=\sqrt{\frac{2EI_3-M^2}{I_2(I_3-I_2)}}
\hbox{sn}\,\tau\cr
&\Omega_3=\sqrt{\frac{M^2-2EI_1}{I_3(I_3-I_1)}} \hbox{dn}\,\tau\,,\cr}
$$
where cn~$\tau$, sn~$\tau$, and dn~$\tau$ are elliptic Jacobi functions
and the dimensionless time $\tau$ is
$$
\tau = t\sqrt{\frac{(I_3-I_2)(M^2-2EI_1)}{I_1I_2I_3}}\,.
$$

Specifying the relative differences $\Delta I_{12}/I_3$, $\Delta I_{23}/I_3$
and expressing the precession phase 
in units of the dimensionless precession period 
$$
\Pi=4\int_0^{\pi/2}\frac{du}{\sqrt{1-k^2\sin^2 u}}
$$
with the parameter $k$ defined as
$$
k^2=\frac{(I_2-I_1)(2EI_3-M^2)}{(I_3-I_2)(M^2-2EI_1)}\,,
$$
we calculate the curves shown in Fig. 2 for the position of the
magnetic pole on the neutron star surface and the angular momentum vector
as explained in the figure caption.

\end{document}